\documentclass[prd,preprint,amsmath,nobibnotes]{revtex4}
\usepackage{bm}
\usepackage{graphicx}

\def\ba{\begin{eqnarray}}
\def\ea{\end{eqnarray}}
\def\be{\begin{equation}}
\def\ee{\end{equation}}
\def\({\left(}
\def\){\right)}
\def\[{\left[}
\def\]{\right]}
\def\<{\left<}
\def\>{\right>}

\newcommand{\labeq}[1] {\label{eq:#1}}
\newcommand{\eqn}[1] {(\ref{eq:#1})}
\newcommand{\labfig}[1] {\label{fig:#1}}
\newcommand{\fig}[1] {\ref{fig:#1}}

\begin{document}

\title{Langevin Analysis of Eternal Inflation}
\date{March 7th, 2005}
\author{Steven Gratton}
\email{S.T.Gratton@damtp.cam.ac.uk}
\author{Neil Turok}
\email{N.G.Turok@damtp.cam.ac.uk}
\affiliation{Department of Applied Mathematics and Theoretical
  Physics,
Centre for Mathematical Sciences, University of Cambridge, Wilberforce
Road, Cambridge. CB3 0WA.  United Kingdom.}

\begin{abstract}

It has been widely claimed that inflation is generically eternal to the
future, even in models where the inflaton potential monotonically
increases away from its minimum. The idea is that quantum fluctuations
allow the field 
to jump uphill, thereby continually revitalizing the inflationary process
in some regions.
In this paper we investigate a simple model of this process,
pertaining to  $\lambda \phi^4$  inflation, in which
analytic progress may be made. We calculate several quantities of interest,
such as the expected number of inflationary efolds, first without
and then with various selection effects. With
no additional weighting, the stochastic noise has little
impact on the total number of inflationary efoldings even if the
inflaton starts with a Planckian energy density.  
A ``rolling'' volume factor, i.e.\ weighting in proportion
to the volume at that time, also leads to a monotonically decreasing
Hubble constant and hence no eternal inflation.
We show how stronger selection effects
including a constraint on the initial and final states and
weighting with the final volume factor can lead to a picture
similar to that usually associated with eternal inflation.

\end{abstract}

\maketitle

\section{Introduction}

If inflation is to provide a satisfactory explanation of the early
universe, it needs both to find a successful microphysical
implementation and to answer the question of why the universe started
out in a high energy inflating state.  The phenomenon of ``eternal
inflation''~\cite{Steinhardt:1983,Vilenkin:1983xq} has been proposed as
a solution of this second
problem~\cite{Linde:1986fc,Linde:1986fd,Linde:1987aa,Guth:2004tw,Vilenkin:2004vx}. 
Even if the inflaton potential monotonically increases away from its
minimum, quantum 
fluctuations allow the inflaton field to jump uphill in some regions
which would then expand exponentially. It is argued that this process,
once started, can allow inflation to continue indefinitely, and that
in all likelihood there was a great deal of inflation in the past of
any observers ``like us''.  The phrase warns that selection effects
are involved, and that these might be important in evaluating the
predictions for various inflationary models~\cite{Aguirre:2004qb}.  At
the simplest level an 
observer ``like us'' could not live in a phase still undergoing
high-energy inflation. So, even if such a phase dominates spacetime,
no observers ``like us'' are there to see it.  Attempts have been made
to implement selection effects with the aid of a volume-weighting, assigning
more weight to larger regions of the universe, which are assumed to
contain a greater number of ``typical'' observers.

The anthropic leanings of the discussion can be disguised by
attempting to rephrase selection effects in a more physical way, for example by
demanding enough inflation to provide superhorizon correlations on the
last scattering surface say. This amounts to the requirement that
there be something like fifty or more inflationary efolds in typical
models (see e.g.~\cite{Tegmark:2004qd}).  But the observed superhorizon
correlations in the universe (including large scale homogeneity and isotropy)
appear to be much stronger than those required for successful galaxy
formation.  If large amounts of inflation are exponentially less
likely than small amounts, one might interpret the existence of
superhorizon correlations in our cosmic microwave background as
evidence \emph{against} inflation.

In any case, it seems important to try and develop calculations of 
conditional probabilities within inflationary models, taking
into account the back-reaction of quantum fluctuations on the
process of inflation itself. 
At first glance, one might think of a slow rolling inflaton field as being
similar to an overdamped harmonic oscillator in the presence
of weak stationary noise.  After a short time such an oscillator ends up at
the bottom of its potential and it only rarely fluctuates appreciably
upwards.  Memory of 
initial conditions is exponentially suppressed, and small fluctuations
away from the minimum are exponentially more common than
larger ones.  If such a model were correct, 
one might expect ``our'' universe to only
have the minimum possible number of efoldings consistent with the
existence of a galaxy say.

As we relate here, taking into account the field dependence of the
Hubble damping and noise leads to a qualitatively different picture.
As we shall discuss, with a change of variable we see that
the system is actually an {\it upside-down}, over-damped
harmonic oscillator, for which there is
no stationary state.

Nevertheless, the system can be studied via a simple Langevin equation
which, for a particular choice of inflaton
potential $V=\lambda \phi^4$ and in the slow-roll approximation, 
is linear and hence exactly soluble.
Even though the dynamical evolution of the system 
is trivial, the quantities we are interested
in, such as the number of inflationary efolds, are highly nonlinear
and nonlocal in time and hence nontrivial to compute. 
Nevertheless, we are able to make 
progress and obtain a number of new results, extending earlier
works on stochastic 
inflation~\cite{
  Starobinsky:1986,Nakao:1988yi,Nambu:1988je,Nambu:1989uf,
  Goncharov:1987ir,Linde:1993xx,Garcia-Bellido:1993wn}.      
The main advantage of the Langevin approach we take over that 
involving the Fokker-Planck
equation (see e.g.~\cite{Linde:1993xx,Vanchurin:1999iv,Aryal:1987vn})
is that
quantities which are non-local in time, such as the number of efolds,
are more easily treated analytically. We are also able to
include
various proposed 
weightings such as volume factors rather straighforwardly.

Before we outline these calculations, however, let us explain 
the physical setup which we believe is approximately described
by the simple stochastic model we employ.

\section{Causal Interpretation of a Hubble Volume during Inflation}

We shall be following a region whose size is the Hubble radius
during inflation, and computing the evolution of the spatially-averaged
field $\phi(t)$ in this region. The justification for focusing on 
just one such Hubble volume is that it 
spans the past light cone
of an observer located far to the future. The spacetime volume
inside this past light cone can be considered as an isolated physical system:
given initial (or final, or mixed initial and final) conditions and a set of 
dynamical laws, its state should be completely describable without
reference to the exterior. 
As long as causality holds it cannot be influenced by anything
outside it.

Consider the past light cone emanating from some point at time $t_1$
in a universe described by a flat FRW metric with scale factor $a$.  
At an earlier time $t$ this light cone encloses a sphere of physical
radius 
\be
r_\textrm{phys} (t) = a(t) \int_{t}^{t_1} dt' /
a(t').
\labeq{one}
\ee  
If $a(t)$ is increasing quasi-exponentially, i.e.\ the Hubble parameter
$H(t) \equiv \dot{a}/{a}$ is positive and only slowly varying with
time, then the integral is dominated by its lower limit and
$r_\textrm{phys} (t)$ becomes approximately equal to $1/H(t)$, the
Hubble radius at time $t$.

So we are interested in describing the evolution of the scalar field
when averaged on the scale of the Hubble radius at that time. We shall employ
a simple stochastic model to describe the scalar field fluctuations
which is standard in discussions of eternal inflation. According to
the model, the scalar field acquires a fluctuation $\delta \phi \sim H$
on a scale of order the Hubble radius, every Hubble time. 
We may understand the
equations as describing the history of the fluctuating 
field as seen in the past light
cone of a point in the far future.

Usually one hears that the physical size of a region that is known to
be inflating increases rapidly.  This statement is only accurate in
the case that the initial region is so large that causal influences
coming from outside the initial region cannot propagate far enough
into it so as to shut inflation down.  The critical size turns out to
be the inflationary Hubble radius. The stochastic model applies to a
region that 
lies on this knife edge: small enough to lie within the past
light cone of a future observer, and 
large enough to 
remain in an inflating state well into the future.
See
Fig.~\fig{conformal}.  

\begin{figure}
\includegraphics[width=12cm]{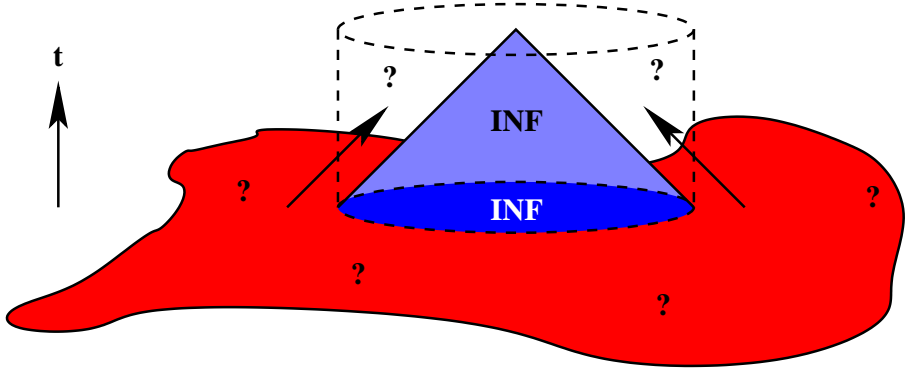}
\caption{
\labfig{conformal} 
Heuristic conformal diagram showing an  inflating patch of critical
size, indicated by the spacetime within the cone. 
The physical radius of the patch at time $t$ is $1/H(t)$; the
physical volume of the patch does not increase exponentially in
time.  
}  
\end{figure}

\section{Background Evolution}

Our starting point for the stochastic approach to inflation will be
the Friedmann and scalar field equations for a flat FRW metric:
\ba
H^2-\frac{8 \pi G}{3}\left( {\frac{1}{2}\dot{\phi}}^2+V \right)&=&0,
\labeq{fe}\\ 
\ddot{\phi}+3 H \dot{\phi}+V_{,\phi}&=&0. \label{eq:sfe} 
\ea
One then considers linearized perturbations about a background
solution. In each Hubble time, new quantum
fluctuations in the scalar field are generated and freeze out
on the scale of the Hubble radius
with an amplitude of order $H$~\cite{Vilenkin:1983xq}.

We model the effects of such fluctuations by adding a stochastic noise
term onto the right-hand-side (RHS) of Eq.~(\ref{eq:sfe}).  We take
the noise to be proportional to delta-function-normalized Gaussian
white noise, $n(t)$, which obeys
\be
\<n(t)\>=0, \qquad \< n(t) n(t')\> = \delta(t-t').
\labeq{noisedef}
\ee
Clearly, $n(t)$ has dimensions of mass to the power one half.
The coefficient may be determined, up to a numerical
coefficient of order unity, by dimensional analysis.  The only
scale entering into the fluctuations is $H$, with dimensions of mass.
Thus we need $H^{5/2} n(t)$ on the RHS
of~\eqn{sfe} for the correct dimensions.  We shall not be concerned
about the numerical coefficient here, but if desired this can be 
determined by normalizing the model to 
a calculated quantum 
correlation function (see e.g.~\cite{Starobinsky:1986}).

There is a long history of attempts to 
to take account of the effects of the fluctuations on the
background evolution in inflation, going back to~\cite{Bardeen:1983qw}
for example.  Ref.~\cite{Vilenkin:1983xq} 
develops a picture of the field evolution over a flat region of its
potential as Brownian motion.  It was later realized that this
motion could also be important for fields with unbounded potentials and
that this effect might be crucial in predicting what an observer might
expect to observe (see~\cite{Linde:1986fc,Linde:1986fd,Linde:1987aa}).
Numerical simulations of this process have been performed
(see e.g.~\cite{Garcia-Bellido:1993wn,Vanchurin:1999iv,Aryal:1987vn}). 

One can often self-consistently make the slow-roll approximation, even
in the presence of noise, which involves neglecting the $\dot{\phi}^2
/2 $ term in~\eqn{fe} and the $\ddot{\phi}$ term in the noisy version
of~\eqn{sfe}, leaving us with 
\ba
3 H \dot{\phi}+V_{,\phi}= H^{5/2} n(t), \labeq{srsfe}
\ea
where $H=H(\phi)=\sqrt{8 \pi G V(\phi)/3}$. From now on we shall
adopt reduced Planck units, setting $8 \pi G=1$.

\section{ Explicit Solution for slow-roll
  $\bm{\lambda \phi^4}$} 

There is a special choice of potential for which~\eqn{srsfe}
simplifies, namely for  $V=\lambda \phi^4$,
as noticed by Nambu~\cite{Nambu:1989uf}.
This is an interesting potential to investigate
in its own right and might reasonably be expected to 
be representative of
other models with simple power law potentials 
such as $m^2 \phi^2$.
After a change of
variable, the Langevin equation becomes linear, with field-independent
noise. Remarkably, the new stochastic variable is just the physical Hubble
radius.

Dropping the second-derivative term, defining
$R=1/ (\sqrt{\lambda/3} \, \phi^{2})$, the physical Hubble radius in the
slow-roll approximation, and introducing
$\alpha=8\sqrt{\lambda /3}$ and $\beta= 2 \sqrt[4]{\lambda /3} /3$
we have
\be \labeq{req}
\dot{R}- \alpha R = \beta n(t),
\ee
where we have changed the sign of the RHS relative to~\eqn{srsfe}.
As introduced above, $n(t)$ is delta-normalized Gaussian white noise with 
zero mean.  We use
angular brackets $\< (...) \>$ to denote the ensemble average of a quantity
$(...)$ over histories of the noise function.  We shall use
double angle brackets $\<\< (...)^q \>\>$ to denote the $q$th cumulant
of the distribution of a quantity $(...)$, and also for
connected correlation functions of products of quantities. 
Note that powers of $\beta$ count the number of times that the noise
enters into any expression.  

Eq.~\eqn{req} describes, as claimed, an over-damped, upside-down
harmonic oscillator with linear noise. It immediately reveals a potential
problem in that 
nothing prevents $R$ from crossing zero for some trajectories. Both
the stochastic model and the underlying field theory of inflation may
be expected to break down there, since the scalar field tends to
infinity. A similar 
pathology occurs in the Fokker-Planck
approach~\cite{Nambu:1989uf,Linde:1993xx}, and
we can try to apply analogous workarounds here where necessary.

If we specify an initial condition $R=r_0$ at $t=0$ say, we have the integral
solution
\be \labeq{rint}
R(t) = r_0 e^{\alpha t} \(1 + \frac{\beta}{r_0} \int_0^{t}
dt_1 e^{-
  \alpha t_1} n(t_1) \).
\ee
Averaging over the noise,  we find
\ba
\<R(t)\>&=&r_0 e^{\alpha t}, \labeq{rav} \\
\<R(t) R(t')\> &=&
r_0^2 e^{\alpha (t+t')}
\( 1 + \frac{\beta^2}{2 \alpha
  r_0^2} \(1 - e^{-2 \alpha \min(t,t')}\) \). \labeq{r2av}
\ea
For the second expectation value we have multiplied two integral
solutions together, taken the ensemble average using the assumed
properties of the noise, given in~\eqn{noisedef}, and then 
performed the integrals over the dummy time variables.

The mean value $\<R(t)\>$ does not involve the noise at all. Changing
variables back
to $\phi$ we see this just represents the classical slow-roll solution
$\phi= \phi_0  
e^{-4\sqrt{\lambda /3} t}$. Because 
the noise is Gaussian, the $R$-distribution is
also Gaussian, with a mean $\mu= r_o e^{\alpha t}$ from~\eqn{rav} and
a variance $\sigma^2 =\beta^2 (e^{2 \alpha t}-1)/ (2 \alpha)$ given by
setting $t'=t$ 
in~\eqn{r2av}.  More generally, one can derive a simple integral
expression for the expectation value of any function $f(R(t))$ of $R$ that
can be expressed as a Fourier integral, by taking the averaging inside
the Fourier integral and writing $\<e^{i k R} \>$ in terms of
cumulants as $e^{ik\<R\> -k^2 \langle \langle R^2 \rangle\rangle /2}$.

Defining $X^2$ as a dimensionless measure, $\sigma^2/ \mu^2$, of the variance
 in $R(t)$:
\ba
X^2 (t)\equiv \frac{\<\< R^2(t)\>\>}{\< R(t)\>^2}= \frac{\beta^2}{2 \alpha
  r_0^2} \(1-e^{-2 \alpha 
  t} \), 
\ea
we see that $R=0$ lies $1/X$ standard deviations to the left
 of $\<R(t)\>$.  Thus if $X$ is small, one can hope that for
 reasonable weightings
 the problems associated with $R=0$ can be neglected.  $X$ is small at
 small times for all starting values
  $r_0$, and remains small for all time if $r_0$ is large enough that
  $X^2_\infty \equiv
  X^2 (\infty) = \beta^2 / (2 \alpha r_0^2)\ll 1 $.  $X^2_\infty$ is a
  quantity with a nice physical meaning: it is, up to a numerical
  coefficient, the ratio of the initial energy density to the Planck
  energy density.  So if we start well below the Planck scale, we do
  not expect subtleties at $R=0$ to be important when following a
  patch forward in time. 

  Strictly speaking, 
one should not trust the theory at all if the density approaches the
  Planck density.  Nevertheless, one can attempt to patch it up 
by imposing 
reflecting~\cite{Nambu:1989uf} or
 absorbing~\cite{Linde:1993xx} boundary conditions at $R=0$.  We
 shall discuss the latter in Sec.~\ref{sec:totalnum} below.  In our
 Langevin treatment these two prescriptions 
correspond respectively to averaging or differencing the probability
densities over
trajectories that start at both $r_0$ and $-r_0$.

\section{Asymptotic Expansion for the Number of Efoldings}

Let us use our integral solution to calculate the expected number
of inflationary efolds
$N(T) =\int_0^{T} dt H(t)$.  If we were following a
comoving region  of a homogeneous inflating universe rather than one
physical Hubble volume in an inhomogeneous universe, $N(T)$
would simply be the number of inflationary efolds that the region had
experienced up to time $T$.  We shall continue to call $N$ the number of
efolds, keeping in mind though that it bears no simple relation
to the physical size of the volume that we are actually tracking
in the calculation. 

Since $H$ is just $1/R$ in our approximation,
\ba
\< N(T) \> \!= \!\int_0^{T}\!\!\! dt \< H(t) \> =\! \int_0^{T} \!\!\!dt
\< \frac{e^{-\alpha t}}{r_0} 
  \(1+I(t) \)^{-1} \>,\labeq{Nexp}
\ea
where
\ba
I(t) \equiv \frac{\beta}{r_0} \int_0^t dt_1 e^{-\alpha t_1}
  n(t_1). \labeq{Idef}
\ea
We now perform a formal Taylor expansion in $I(t)$ of the reciprocal
inside~\eqn{Nexp}.  The justification for this procedure is 
that as long as the noise term can be treated as ``small'', an expansion
in the noise makes sense.

We can take the expectation value of powers of $I$ term by term making use
of the fact that the noise is assumed to be Gaussian so that all
higher-order correlation functions of it can be expressed in terms of
products of its two-point function via Wick's theorem.  It thus turns out
that $\< I^q (t) \>$ vanishes if $q$ is odd, and equals
$(2p-1)!! X^{2p}(t)$ if $q=2p$ is even, with $X(t)$ as
defined above and $(2p-1)!!= (2p-1)(2p-3)\ldots 1$.  
Thus
\ba
\< H (t) \> = \frac{e^{-\alpha t}}{r_0} \( 1 + X^2 (t) +3 X^4(t) +
\ldots + (2p-1)!! 
X^{2p} (t)+\ldots \).\labeq{Hinf}
\ea
Finally, we can perform the integral over $t$ term-by-term to obtain
a series for $\< N(T)\>$.  

This procedure works particularly neatly if we take $t \rightarrow
\infty$, since then the time integrals are straightforward, using
$\int_0^\infty dy 
e^{-y} (1-e^{-2 y})^p = 2^p p!/(2p+1)!!$.
With $N_\infty \equiv N(\infty)$, one obtains:
\ba
\< N_\infty \> = N_{\text{cl}} \left( 1+ 1\cdot \frac{2}{3} \cdot
  X^2_\infty+ 3 \cdot   
\frac{16}{30} \cdot X^4_\infty + \ldots + \frac{ 2^p p!}{2p+1}
X^{2p}_\infty +\ldots\), \labeq{Ninf}
\ea
where $N_{\textrm{cl}}=\frac{1}{\alpha
  r_0}$ is the ``classical'' number 
of efolds expected in the slow-roll approximation at late times in the
absence of 
noise.    

Unfortunately the two series expansions~\eqn{Hinf} and~\eqn{Ninf} diverge.  
We shall investigate this divergence in more detail in the next section.
For now, let us understand these series as asymptotic
expansions in terms of $X^2_\infty$, the initial energy density.
This is plausible since the Taylor expansion of $(1+I(t))^{-1}$ inside
the integral cannot be valid for all
histories of the noise.  But if the noise is weak
enough we do not expect this inaccuracy to be significant. 
An analogous example is the calculation of the 
expectation value of $(1+\epsilon x^2)^{-1}$ or $(1-\epsilon x)^{-1}$
say for
a Gaussian distribution with zero mean and unit variance in the case
that $\epsilon \ll 1$, for which the perturbative expansion,
although only asymptotic, is still extremely accurate.  

Our second illustration $(1-\epsilon x)^{-1}$ above is instructive in
suggesting a formal procedure to account for effects associated with
$R=0$.  Here we have a divergence in the calculation
of the expectation value, at $x=-1/\epsilon$, far into the
tail of the distribution.  If we don't trust the detailed form of the
distribution for large negative $x$, we may feel that such a
divergence is unphysical and irrelevant.  A natural mathematical way
of removing such a divergence, and allowing us nevertheless to
continue to use the 
Gaussian form, is to calculate the
expectation value of principal values (denoted $\textrm{P.V.}$) of
linearly-divergent quantities rather than the quantities themselves.
In the stochastic model this prescription turns out to be related to 
imposing absorbing boundary conditions at $R=0$.

After these comments, let us see what we can glean from
Eq.~\eqn{Ninf}.  We see that fluctuations increase the expected number
of efolds that a patch undergoes, but only by a small multiplicative
fraction if the energy density starts low compared to the Planck density.
Significant corrections to the classical result only occur at such
high energy densities that we can have no confidence in the applicability
of the theory.

\section{The Total Number of Efoldings}
\label{sec:totalnum}

We would like to investigate in greater depth the divergence of our series
for $\< H(t) \>$ and $\< N (T) \>$. 
In the Fokker-Planck approach, a natural way to regulate the problem
at $R=0$ is 
to impose the ``absorbing'' boundary condition, that the probability
is zero at $R=0$~\cite{Linde:1993xx}.  This can be done for a
Gaussian probability distribution by reflecting it around $R=0$ and
subtracting the reflected copy off the original one.  This new
probability distribution 
must then be renormalized.  In the Langevin approach this
procedure corresponds to ignoring any paths that reach $R=0$.
One could alternatively impose ``reflecting'' boundary conditions for
which the gradient of the probability distribution at $R=0$ is zero,
or indeed consider a mixture of the two conditions.  
Only the pure absorbing condition ensures that the probability
vanishes at $R=0$, and for definiteness we shall focus on this case here.

From our Langevin approach, we know that before the boundary condition
is imposed $R$ is Gaussian-distributed, with mean $\mu$ and variance
$\sigma^2$ given above.
Applying the
reflect-and-renormalize procedure outlined 
above, the expectation value of some function $f(R)$ is given by
\ba
\< f(R) \>_{\text{abs}} \negmedspace&=& \negmedspace
\frac{ \int_{0^+}^{\infty} dx  f(x) \(
  e^{-\frac{(x-\mu)^2}{2 \sigma^2}}- e^{-\frac{(x+\mu)^2 }{2 \sigma^2}} \)}
{\int_{0^+}^{\infty} dx 
  \(
  e^{-\frac{(x-\mu)^2 }{2 \sigma^2}}- e^{-\frac{(x+\mu)^2 }{2
      \sigma^2}} \)}.
\labeq{rabs}
\ea
(The subscript ``abs'' is for absorbing.)  We extend the range of
$f$ from the positive reals to the entire real line by demanding that
$f$ is odd.  Then we let $x \rightarrow -x$ in the second term of the
upper integral, and divide top and bottom by $\sqrt{2 \pi \sigma^2}$.
  The numerator is then the expectation value of the P.V. of $f$ 
with respect to the Gaussian we started with over the entire real
line.  The denominator is the error function $\text{erf} (\mu / (\sqrt{2}
\sigma))= \text{erf} ({1/(\sqrt{2} X(t)))}$.  

Since $H(t) =1/R(t)$ we can apply this to work out $\<H(t)\>_{\text{abs}}$:
\ba
\< H(t) \>_{\text{abs}}=
\frac{ \< \text{P.V.} \( \frac{1}{R(t)} \)  \>}{\text{erf}\(
  \frac{1}{\sqrt{2} X(t)} \) }.
\ea
To make contact with our asymptotic expansion above, we write $
\text{P.V.} \( \frac{1}{R(t)} \) $ as a Fourier integral and then take
its Gaussian expectation value as discussed earlier.  This gives the
numerator as the integral 
\ba
\< \text{P.V.} \( \frac{1}{R(t)} \)  \> =
\frac{1}{\< R(t) \>} \cdot \frac{1}{X(t)} \int_0^{\infty} dk
\sin\(\frac{k}{X(t)}\) e^{-k^2/2}. 
\ea
If we then continually integrate by parts on the $\sin$ we generate
the asymptotic expansion~\eqn{Hinf}.  

We still have to consider the denominator.  Now
$\text{erf} (1/(\sqrt{2}X(t))=1-\text{erfc}(1/(\sqrt{2} X(t)))$, and
for small $X(t)$ $\text{erfc} (1/(\sqrt{2} X(t)) \sim \sqrt{2/\pi}
X(t) e^{-1/(2 X^2 (t))}$ (see e.g~\cite{Abramowitz}) which is exponentially
  small.  So the denominator itself is exponentially close to one, and
  the entire expression is accurately approximated by the asymptotic
  expansion~\eqn{Hinf}.

Thus we see that effects coming from imposing boundary conditions at
$R=0$ are negligible to the extent that $X(t)$ is small.

This condition at $R=0$ does allow us to go on to investigate the
large $X$ regime also.  We may rewrite~\eqn{rabs}, applied to
$H(t)=1/R(t)$, as
\ba
\< H(t) \>_{\text{abs}} = \frac{\int_0^\infty \frac{dx}{x} W(x)}  
{\int_0^\infty dx W(x)}
\ea
where
\ba
W(x)\equiv 
\sinh \left(\frac{x \mu}{\sigma^2} \right) e^{-\frac{x^2}{2\sigma^2}}.
\ea
By rescaling the integration 
variable, one finds that the expected Hubble constant can be expressed
in the form
\begin{equation} 
\< H(t) \>_{\text{abs}} = \frac{\mu}{\sigma^2} F(\sigma^2 / \mu^2).
\end{equation}
One finds the asymptotic  
behaviour: 
\ba 
F(z) \sim z, \quad &\text{for~}z \ll 1; \\ 
F(z) \sim  \sqrt{(\pi z)/2}, \quad &\text{for~}z \gg 1.
\ea
Using this, one estimates the expected total number of efoldings
$\<N_\infty\>_\text{abs}= \int_0^\infty dt  \< H(t) \>_{\text{abs}}$ as:
\begin{equation}
 \<N_\infty\>_\text{abs} \sim  \frac{1}{\alpha r_0} =
 \frac{\phi_0^2}{8}, \quad \text{for~} r_0 \gg 1, 
\end{equation}
which is just the standard classical result, and 
\begin{equation}
\<N_\infty\>_\text{abs} \sim  \Big( {\frac{\beta^2}{ 2
 \alpha}}\Big)^{-1/2} \left(\frac{\pi}{2}\right)^{3/ 2} \frac{1}{\alpha},  
 \quad \text{for~} r_0 \ll 1,
\end{equation}
which up to a numerical constant is $\lambda^{-1/2}$, 
just the standard classical slow-roll
result for inflation starting at the Planck density. Hence we 
conclude that quantum fluctuations, treated as stochastic noise,
do not qualitatively alter the expected number of inflationary
efoldings beyond the classical result, for initial densities right
up to the Planck density.

\section{``Eternal Inflation'' from Strong Selection}

Imagine we know that at two times the scalar field was well up
the hill within the Hubble volume we are tracking.
What can we say about the likely value of the field
at intermediate times? Did the scalar field roll downhill and
then fluctuate back up, or did it fluctuate up and then roll
down to its prescribed final value?

There are two equivalent ways of calculating such an effect.
One is to use Bayes'
theorem explicitly.  The other to 
exploit the fact that the noise is Gaussian, which means that the full
probability distribution for multiple events is given by
exponentiating their covariance matrix (this method cannot be used if
the boundary conditions at $R=0$ are important).  We shall present the former
method here.

Let us take $0<t<T$. We denote by $p_{1|1} (r_t| r_0) dr $ the 
probability that at time $t$, $R$ lies between $r_t$ and $ r_t+dr$
given that $R(0)=r_0$; and
by $p_{1|2} (r_t|r_T,r_0) dr$ the probability that at time  $t$, $R$ lies
between $r_t$ and $ r_t+dr$ given that $R(T)=r_T$ and $R(0)=r_0$. 

Then, by Bayes' theorem,  
\ba
p_{1|2} (r_t|r_T,r_0)=\frac{p_{1|1} (r_t|r_0) p_{1|2} (r_T
  |r_t,r_0)}{p_{1|1} (r_T|r_0)} \labeq{bayes}
\ea
and we know what all the terms on the RHS are, since $ p_{1|2} (r_T
  |r_t,r_0)= p_{1|1} (r_T|r_t)$.

If we neglect $R=0$ effects, the probability distributions are Gaussian
and we can use our formulae~\eqn{rav}
and~\eqn{r2av} to evaluate the means and variances for the RHS.  With
the constraints on $R$ at $0$ and $T$ we find that at the intermediate
time $t$, $R$ is again Gaussian-distributed with a $t$-dependent mean of   
\ba
\<R(t)\>_{\text{con}}=\frac{r_0 \sinh \alpha (T-t) + r_T \sinh \alpha t}{\sinh
  \alpha T}.
\ea
(The subscript ``con'' is for constrained.) In the Appendix we discuss
an interesting formal property of 
$\<R(t)\>_{\text{con}}$, which is that it 
obeys a second order differential 
equation related to the original first order equation (\ref{eq:req})
in a simple way. 

One can also easily
find the variance in $R$ by looking at the 
coefficient of the term quadratic in $r_t$ in the exponent on the RHS
of~\eqn{bayes}.  It turns out to be:
\ba
\<\<R^2(t)\>\>_{\text{con}}=\frac{\beta^2}{2 \alpha} \frac{2 \sinh \alpha (T-t)
  \sinh \alpha t }{\sinh \alpha T}.
\ea
Interestingly, this is independent of $r_0$ and $r_T$, and attains its
maximum value of $(\beta^2 / 2 \alpha) \tanh (\alpha T /2)$ at the
midpoint $t=T/2$.  This is less than $\beta^2 / 2 \alpha $, so the
standard deviation 
in $R$ here is 
always bounded by a constant of order unity. 
As long as
$\<R (t) \>_{\text{con}}$ is always 
significantly larger than unity, we expect that we may neglect 
effects from $R=0$. 

Furthermore, if the minimum of $\<R(t)\>_{\text{con}}$ is much less than
$r_T$, then relatively close to $T$ $\<R(t)\>_{\text{con}}$ behaves like
$r_T e^{\alpha (t-T)}$, independent of $r_0$.  This is the same as the
noise-free slow-roll solution which passes through $r_T$ at time $T$,
and which in the far past would approach arbitrarily small $R$ and
thus large $\phi$.

For example, if we take $r_T=r_0$, $\<R (t) \>_{\text{con}}$ takes a minimum
value of $r_0 / \cosh (\alpha T/2)$ at $t=T/2$.  Demanding that the
minimum value is greater than the standard deviation at that time puts
a condition on 
the values of $T$ for which we may reliably neglect effects associated 
with the $R=0$ boundary: this condition reads 
$\sinh \alpha T <2/ X^2_\infty$.  Since $ X^2_\infty$ is
typically small if the field starts low down the potential, we see
that as long as $T \lesssim \ln (4/ X^2_\infty) / \alpha$ boundary
effects from $R=0$ should be unimportant.
As long as this is so, we can now answer the question posed at the
beginning of this section. 
If the field is known to be at the same place on the potential at
times $0$ and $T$, the Hubble radius lies at smaller
values in the interim, corresponding to the field being further up the
potential.  See Fig.~\fig{contraj}.

\begin{figure}
\includegraphics[width=12cm,height=10cm]{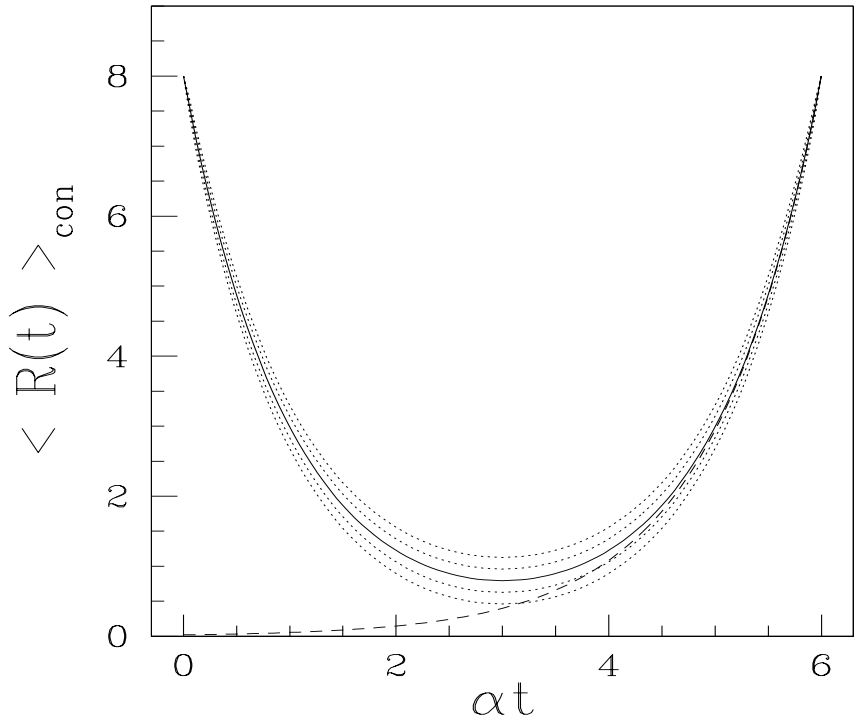}
\caption{
\labfig{contraj} 
A plot showing the expectation of $R=1/(\sqrt{\lambda/3} \phi^2)$
against time for paths constrained to take specified values at two
different times.  In the example shown, 
$R=8$ at $\alpha 
t=0$ and $\alpha t= 6$ (solid line).  Also indicated are one and two sigma
contours for the trajectories (dotted lines), and the noise-free
solution for $R$ that passes through $R=8$ at $\alpha t=6$ (dashed
line).  We have set
$\beta^2 / 2 \alpha = 1/36$ in making this plot.
}  
\end{figure}

So we have shown that if inflation has lasted a 
reasonably long time
(i.e.\ we know that the field was displaced well up its potential at
two widely separated times), the field was likely to have rolled up
the hill to higher field values, and then turned round and rolled back
down, approaching the standard slow-roll solution.
This behaviour shares many of the characteristics that are commonly
associated with the phrase ``eternal inflation''.

As an example, imagine we select paths in which the scalar
field takes a value of order the Planck mass at two widely separated
times.  Recall that this is the value for which the slow-roll
approximation fails, near the end of inflation.
The above calculation shows that the expected trajectory is one in which
the field first runs uphill to very large values before
rolling down along a nearly classical slow-roll path to the final
value. For such paths, one would expect
the usual predictions for the observable density perturbations.

However, by demanding that $T$ is large we have imposed a very strong
selection of paths; the vast majority of paths run downhill.  Now let
us consider weakening the selection by reducing the time $T$ between
our initial and final states. Even if we insist upon fifty efoldings
of inflation, the likely trajectory is very different from the
classical slow-roll one usually considered.  In standard inflation
where one ignores the effect of quantum fluctuations on the
background, the density perturbation amplitude is approximately
$H^2/\dot{\phi}$.  However, in the selected ensemble the velocity
$\dot{\phi}$ of the scalar field vanishes at an intermediate time.
This would lead one to suspect that for the selected ensemble, density
perturbations which exited the Hubble radius around the time $T/2$
would be far larger than those in standard inflation. Unless $T$ is
chosen to give substantially more than fifty efoldings, the normal
predictions of inflationary perturbations will not be obtained.

To conclude this section, 
we have seen how selection effects can indeed make a scalar field
roll uphill before rolling roll back along a near-classical slow-roll path.
This is the same type of behaviour as that sought from eternal inflation,
but we have obtained it through an imposed ad-hoc selection, rather than 
from volume weighting, the effect to which it is normally attributed. 
We now turn to an investigation
of volume weighting to see whether this can produce a similar effect.

\section{Eternal Inflation as a Volume Effect}

The usual argument for inflation being eternal to the future with an
inflaton field with an unbounded potential is summarized and criticized
in~\cite{Hawking:2003bf}.  The main idea is that regions of space in
which the field fluctuates higher inflate more rapidly, and thus after
a long amount of time the majority of the physical volume is dominated
by inflating regions at the Planck density.  Occasional regions
fluctuate downwards and inflation stops, and we are expected to live
in one of these regions.  The main criticism is that this line of
argument is not gauge invariant, i.e.\ comparisons of volumes depend on
the choice of time-slicing.

We argued above that the stochastic equation really only follows the
evolution of the field inside one physical Hubble volume.  The common
supposition is that it rather describes the average behaviour of the
field in a fixed \textit{comoving} volume.  In Fig.~\fig{conformal},
this corresponds to assuming that the field on the constant-time slice
through the entire conformal cylinder is the same as it is on the
piece of the slice inside the cone.  In this case, one may then argue
that expectation values of various quantities evaluated at time $t$
should be calculated by weighting each trajectory by its volume factor
$e^{3 N(t)}$. In this section, we shall explore the consequences of
this suggestion.

We thus define the ``volume-weighted'' average of some quantity
$(...)$ at
time $t$ via:
\ba
\<(...)\>_{\mu=3,t} \equiv \frac{\< (...) e^{3 N(t)} \>}{\< e^{3
  N(t)}\>}.\labeq{voldef}
\ea
(The reasoning behind the subscript will become clear shortly.)

Let us attempt to calculate the volume-weighted number of efolds $
\<N(t)\>_{\mu=3,t}$.  This is an interesting
example because the usual claim (see e.g.~\cite{Guth:2004tw}) is that
if $\phi_0$ 
is above a critical value of order $\lambda^{-1/6}$ then the physical
volume of space that is inflating should increase without limit,
corresponding to eternal inflation.  Thus we would expect to see a
qualitative change in the behaviour of $ \<N(t)\>_{\mu=3,t}$ around
this critical starting value.

We start by introducing the formal generating function $W_t (\mu) =
  \ln  \< e^{ \mu
  N(t)} \>$, where $\mu$ is a number.  Differentiating this with
  respect to $\mu$ 
  and then setting $\mu=3$ gives 
  us the volume-weighted number of efolds, $\<N(t)\>_{\mu=3,t}$ (hence
  the subscript in~\eqn{voldef}).  We
  may think of $W_t (\mu)$    
  as
\ba
\sum_{n=1}^\infty \frac{\mu^n}{n!} \< \< N(t)^n \> \>.
\ea 
So we can express $\< N(t) \>_{\mu=3,t}$ in terms of the connected
  moments or cumulants of $N$.  The $n$th cumulant of $N$ can be
  obtained as usual from the regular moments of $N$, which can be
  calculated order by order in the noise.  (Such a procedure is
  described in more detail in the following section where it is needed
  for a wider-ranging calculation.) 

In the limit $t\rightarrow \infty$, it turns out that the $n$th
cumulant goes like $N_{\text{cl}}^n$ times a power series in $X^2_\infty$
starting at the $(n-1)$th power of $X^2_\infty$.  So to order
$X^4_\infty$ we find 
\ba
\<\<N_\infty\>\>&=&N_\text{cl} \(1+ \frac{2}{3} X^2_\infty +
\frac{8}{5} X^4_\infty+\ldots \) \\
\<\<N^2_\infty\>\>&=&N_\text{cl}^2 \(\frac{1}{2} X^2_\infty +
3 X^4_\infty+\ldots \) \\ 
\<\<N^3_\infty\>\>&=&N_\text{cl}^3 \(\frac{12}{7} X^4_\infty +\ldots \)
\ea
with all higher cumulants higher order in $X^2$.
Following the aforementioned procedure and collecting terms in $X^2$, we find
\ba
\<N_\infty\>_{\mu=3, \infty} = N_\text{cl} \Bigg\{ 1+ \(\frac{3}{2}
  N_\text{cl}+ 
\frac{2}{3} \)  X^2_\infty + \( \frac{54}{7} N_\text{cl}^2+ 9
N_\text{cl}+\frac{8}{5} \) X^4_\infty + O\( X^6_\infty \) \Bigg\}.
\ea
Thus we obtain the fractional correction to the classical number of
efolds $N_{\mathrm{cl}}$ to fourth order in the noise.  One sees that
for $N_{\mathrm{cl}} 
\gg 1$ (but still $X^2_\infty \ll 1$), we in fact seem to have an expansion in
$N_{\mathrm{cl}} X_\infty^2$ rather than in just $X_\infty^2$.  So
our formula gives only a small fractional correction to the number of
efolds as long as $N_{\mathrm{cl}} X_\infty^2 \ll 1$.

But  $N_{\mathrm{cl}} X_\infty^2 = \beta^2 / (2 \alpha^2 r_0^3) \sim
\lambda \phi_0^6$ 
and so we see that when $\phi_0 \gtrsim \lambda^{-1/6}$ the
volume-weighted number of efolds begins to change substantially from
the noise-free result.  This is just the criterion mentioned above
that is usually given for the onset of eternal inflation.

So our method has established some contact with the usual approach to
eternal inflation.  However, we caution that the assumption that one
should volume-weight is not yet (in our view) 
well-founded, since the dynamical
equations being used are only following
the value of the field in one Hubble volume.

\section{A closer investigation of volume-weighting}

In the above section we introduced volume-weighted averages of
quantities defined at a given time, where the volume-weighting was
performed at that time.  We used this to calculate the volume-weighted
number of efolds at a given time $t$, and then considered taking the
limit $t\rightarrow \infty$. 

In this section, we shall consider two sorts of volume weighting.
If one believes in volume-weighting as a physical spatial
averaging, then to compute observables at time $t$ one 
might want to weight with the volume at that time. On the other
hand, if one thinks of volume-weighting as a selection effect for
``typical'' observers, one might
want to weight with the ``final'' volume, at some much later time.
To cover both cases, we shall compute
the Hubble
parameter as a function of time, $H(t_a)$, weighted by the volume at a
later time $t>t_a$.  Taking the limit $t_a \rightarrow t$, we obtain the
trajectory of 
the Hubble parameter with a rolling volume weighting. If instead 
$t$ is taken to $\infty$, we obtain the 
expected history of the Hubble parameter when selected by
final volume.  This gives what a ``typical'' late-time
volume-weighted observer should
expect to see in his/her past.  (Note that our treatment in the previous
section took the ``spatial averaging'' point of view but focused only on
the final number of efoldings.) 

To calculate such averages involving the Hubble parameter, one needs
to generalize the generating function introduced above, promoting
$\mu$ from a number to a function of time.  Thus one defines the
generating functional $W_t [\mu]$ 
\ba
W_t [\mu] = \ln \< e^{M_t [\mu]} \>,
\ea
where
\ba
M_t [\mu] \equiv \int_0^t dt_1 \mu(t_1) H(t_1).
\ea
Now by functionally differentiating with respect to $\mu(t_a)$, and then
setting $\mu=3$ for all time, one obtains volume-weighted 
cumulants involving $H(t_a)$.  Different choices for $\mu $ correspond
to more general forms of weighting, and we may denote expectation values
with respect to them by $\<\< \ldots\>\>_{\mu(t'),t}$ should we
wish to use them.  Note that $\mu=0$ corresponds to the natural
weighting.  It can be useful to take $\mu$ to be an arbitrary
constant at intermediate stages of the calculation, and only set
$\mu=3$ at the very end.  Then one can easily identify any effects
coming from the volume weighting.

In order to proceed perturbatively in the noise, we again expand $W$ in
terms of the regular (i.e.\ non volume-weighted) cumulants of
$M[\mu]$:
\ba
W_t [\mu] =  \sum_{n=1}^\infty \frac{1}{n!} \< \< M^n_t[\mu] \> \>.
\ea
These cumulants involve multiple integrals $\int dt_1 \ldots \int dt_j$
over products of the $H$s.
Each $H$ is written as:
\ba
H (t_i) = \frac{e^{- \alpha t_i}}{r_0} \( 1 - I_i + I_i^2 - I_i^3 + I_i^4
\ldots \) 
\ea
where $I_i$ is a shorthand for $I(t_i)$ as defined in~\eqn{Idef}.  We
use Wick's theorem to express higher-order moments of $I$ in terms of
the two-point function $\< I_1 I_2 \>$, which is given by
\ba
 \< I_1 I_2 \>= X_\infty^2 \( 1- e^{-2 \alpha \min(t_1, t_2)}\).
\ea
Each power of $X_\infty$ counts an order of the noise.  As in the
previous section, it turns out that the $n$th cumulant is of order
$2(n-1)$ in the noise.  

Let us now use the above machinery to calculate $\< H(t_a)
\>_{\mu=3,t}$ to second order in the noise.  We need the first and second
cumulants of $M$, each evaluated to second order in the noise.
Functionally differentiating the former leaves us with $\< H(t_a)
\>$, or
\ba
\frac{e^{- \alpha t_i}}{r_0} \( 1 +  X_\infty^2 \( 1- e^{-2 \alpha t_a}
  \) \).
\ea 
Functionally differentiating the latter cumulant yields $ 2 \int_0^t
dt_1 \mu(t_1) \<\< H(t_1) 
H(t_a) \>\>$, or
\ba
2 \int_0^t dt_1 \mu(t_1) \frac{e^{-\alpha t_1}}{r_0} \frac{e^{-\alpha
      t_a}}{r_0} X_\infty^2 \( 1- e^{-2 \alpha \min(t_1, t_a)} \).
\ea
This can be evaluated assuming $\mu$ to be constant in time.  
Utilizing these results and setting $\mu=3$ leads to 
\ba
\<\< H(t_a)\>\>_{\mu=3,t}&=&
\frac{e^{-\alpha t_a}}{r_0} 
\Bigg\{1+X_\infty^2 \(1-e^{-2 \alpha t_a}\) \nonumber \\
&+&3 N_\text{cl} X_\infty^2 \Big( 1-e^{-\alpha t} -\frac{1-e^{- 3 \alpha
  t_a}}{3} - e^{-2 \alpha t_a} 
  \( e^{- \alpha t_a } - e^{-\alpha t } \) \Big) \Bigg\}. 
\ea

As discussed above, we are interested in two cases. The first,
involves taking $t_a \rightarrow t$. This choice implements
a ``rolling volume weighting'', where at any time the volume-weighted
average is performed using the volume at that time.  
This choice would correspond
to what one might usually consider to be a spatial average 
over physical volume. We obtain 
\ba
\<\< H(t)\>\>_{\mu=3,t}=
\frac{e^{-\alpha t}}{r_0} \Bigg\{1+X_\infty^2 \(1-e^{-2 \alpha t}\)
+3 N_\text{cl} X_\infty^2 \(1-e^{-\alpha t} - \frac{1-e^{- 3 \alpha
    t}}{3}\) \Bigg\}. 
\ea
(This result can also be obtained by calculating
$\<N(t)\>_{\mu=3,t}$ as discussed in the previous section to
second order in the noise and then differentiating with respect to
$t$.)   
A Taylor expansion of the correction term coming from the
volume-weighting starts 
at $O((\alpha t)^2)$.  This is to be compared with the $O(\alpha
t)$ variation coming from the expansion of the $e^{-\alpha t}$
``classical rolling'' term.  One (initial) Hubble time is $t_H=1/H_0=r_0$, so
$\alpha t_H= \alpha r_0=1/N_\text{cl}$, which is small if the
classical number of efoldings is large.  Thus as long as one is
interested in times up to of order a few Hubble times from the start,
rolling volume-weighting does very little even if $ N_\text{cl}
X_\infty^2 \sim
\lambda \phi_0^6 \gg 1$.   One only starts getting
corrections of 
order $ N_\text{cl} X_\infty^2 $ when $\alpha t \sim 1$, that is at 
times for which, without noise,
a substantial fraction of the total number $N_\text{cl}$ of
efolds would have occurred.
We have
checked that including the fourth 
order term does not affect this conclusion, and strongly suspect that
neither will further higher order terms. Hence we conclude that
a rolling volume weighting leads to a monotonically decreasing
Hubble constant, and will not produce the uphill motion 
required by the eternal inflation picture. 

Now let us consider the second limit.  This is to take $t \rightarrow
\infty$.  Thus we are weighting trajectories by their final volume.  As
discussed above, this in
some sense corresponds to the history of the field that a physical
observer at late times might expect to see in his/her past light cone,
assuming that physical observers are evenly distributed over the final
three-volume.  We obtain:
\ba
\< H(t_a) \>_{\mu=3,\infty}=
\frac{e^{-\alpha t_a}}{r_0} \Big\{ 1+ X_\infty^2 \(1-e^{-2 \alpha
  t_a}\) + 2 N_\text{cl} X_\infty^2 \(1-e^{-3 
  \alpha t_a} \) \Big\}.
\ea
Now the volume-weighting term has a Taylor expansion that starts at
$O(\alpha t_a)$, the same order as that from the ``classical''
exponential.  So now for large enough $N_\text{cl}$, we can have the
expectation value of the Hubble parameter increase with time.  The
requirement is that $6 N_\text{cl} X_\infty^2 > 1$, or $\phi_0 \gtrsim
\lambda^{-1/6}$, again the criterion for eternal
inflation. By $\alpha t_a \sim 1$, the fractional correction to the
noise-free result is becoming large and we do not trust our result in
the details.  Of course, if the noise is very important we should not
overly trust the $t\rightarrow \infty$ limit in the first place.

We can investigate the spread in the final-volume-weighted trajectories by
considering the ratio $ \< \< H^2 (t_a) \> \>_{\mu=3, \infty} / \<
H(t_a)\>^2_{\mu=3, \infty}$.  The lowest-order term in the numerator
is already $O(X_\infty^2)$, so we must calculate the numerator to
fourth order to 
obtain the fractional change in the ratio due to volume-averaging to
second order.  Thus we need the third cumulant of $H$ for the
numerator, but the 
calculation can be performed.  Taking $\alpha t_a \rightarrow \infty$
to obtain the late-time behaviour and dropping terms down by
$N_\text{cl}$ we find
\ba
\frac {\< \< H^2 (\infty) \> \>_{\mu=3, \infty} }{\<
H(\infty)\>^2_{\mu=3, \infty}} \approx
X_\infty^2 \(1+ \frac{26}{5} N_\text{cl} X_\infty^2 \).
\ea
Thus we see that in the regime of eternal inflation the noise is
playing an important role.  
One might be tempted to conclude that for  $\phi_0 \gtrsim
\lambda^{-1/6}$ the volume-weighted system becomes ``noise-dominated''
and a ``non-perturbative'' way of treating the noise is called for.

\section{Conclusions}

We have seen that $\lambda \phi^4$ inflation with quantum jumps may be
modelled as an upside-down, overdamped 
harmonic oscillator with noise.  The critical role that selection
effects have on expectations for the field's observed history has been
clearly demonstrated.   
In the absence of any selection
effects, we have seen that the presence of stochastic noise 
hardly alters the motion of the inflaton field downhill: the
total number of inflationary efolds is essentially the same
as the classical result. Through strong selection one can 
obtain trajectories which travel uphill: 
if the field is constrained
to lie at a particular inflating value at two widely separated times
we have shown that it is likely in the
interim to be at higher field values.  

We also investigated volume-weighting in our approach. With
a rolling weighting, such as might be attributed to averaging
over physical volume, we find no significant effect on
the Hubble parameter's evolution until late
times, for which inflation
would classically have completed a substantial fraction of its total
number of efolds.
If, alternatively, histories are weighted by the {\it final} volume,
qualitative 
changes in the mean trajectory of the system can occur within a few
Hubble times of the start. However in this situation,
it is not clear whether the perturbative treatment is valid.

Our techniques might also be applicable for more general forms of
averaging.  For example, one might consider the possibility of
weighting by 4-volume.  This would be the right thing to do if
observers spring into being from inflating regions in a stochastic
manner consistent with Poisson statistics (say by bubble nucleation),
and that the appearance of a region supporting physical observers has
a negligible effect on the background evolution (unlike bubble
nucleation).  If the 4-volume were finite, the result would be
gauge-invariant and ambiguity-free also.  The generating function
could then be $\ln \< \int_0^\infty dt_1 e^{\mu N(t_1) }\>$, and
differentiating this with respect to $\mu$ and then setting $\mu=3$
would give the 4-volume-weighted number of efolds.  The generating
function could be calculated order-by-order in the noise using our
methods.  As we have explained, however, we are not convinced that any
particular form of volume-weighting is physically correct, since the
stochastic equation being used only describes the time evolution of a
single physical Hubble volume.

There are several possible extensions of the work reported here.
Numerical simulations could be used to extend the perturbative treatment
we have given and to check the generality of
the behaviour found here for other potentials. It
would be interesting to know whether inflation models  with
much flatter potentials 
exhibit
the same qualitative behaviour.
It would also be interesting to extend the model to allow the inflaton
field to jump back up from its minimum into an inflating regime,
a situation referred to by Garriga and Vilenkin as a ``recycling 
universe'' \cite{Garriga:1997ef}  (see
also~\cite{Dutta:2005gt}). In this situation one might hope to 
find a steady state solution, although the typical
bout of inflation would involve very small numbers of efolds 
and so such a model would be 
observationally disfavoured.

\begin{acknowledgments}
  
We particularly thank Anthony Aguirre for many useful discussions
and helpful comments on a draft of this paper, and also thank Andy
Albrecht, Misao Sasaki and Max Tegmark for useful conversations.

\end{acknowledgments}

\appendix*

\section{A property of $\bm{ \<R(t)\>_{\text{con}}}$}

In this Appendix we discuss an interesting fact about 
the constrained expectation $\<R(t)\>_{\text{con}}$, 
which is that it satisfies the {\it second order} deterministic differential
equation
\be 
\(\frac{d}{dt} -\alpha\) \(\frac{d}{dt} +\alpha\) \<R(t)\>_{\text{con}} =0,
\labeq{funny}
\ee
where the operator appearing is the product of the original 
operator in the equation of motion, (\ref{eq:req}), and its
time reverse. That this should be so is in fact 
a general consequence of time-translation invariance
of the operator appearing in the original equation 
of motion, (\ref{eq:req}), and of the noise ensemble. 
The point is that the constrained
expectation $\<R(t)\>_{\text{con}}$ is proportional to
the correlator $\<R(t) R(T)\>$, as may be seen by writing 
the joint (Gaussian) probability distribution for $R(t)$
and $R(T)$. To show the correlator obeys the stated equation 
(\ref{eq:funny}), note first that
\be
\(\frac{d}{dt} -\alpha\)\(\frac{d}{dT} -\alpha\)\<R(t) R(T)\> =0
\labeq{funny1}
\ee
for all $t\neq T$, since from the equation of motion (\ref{eq:req})
this equals $\beta^2 \<n(t) n(T)\>$, which is zero at unequal times. 
Now, we can write the left hand side 
as $(\frac{d}{dT} -\alpha)$ acting 
upon $\beta \<n(t) R(T)\>$. But $R(T)$ is a sum of a particular
integral which does not correlate with the noise, and
an integral $\beta \int_0^T dt' G(T,t') n(t')$, where $G(T,t')$ is the Green
function of the original operator. So $\<n(t) R(T)\>$
is just $\beta G(T,t)$. Due to time translation invariance, 
the latter is a function only of the time difference $t-T$.
Therefore, we may replace $\frac{d}{dT}$ in (\ref{eq:funny1}) with
$-\frac{d}{dt}$, obtaining (\ref{eq:funny}). 
The arguments we have used are rather general: it is clear for example
that constrained expectation values related to a second order
stochastic differential equation would obey a fourth order
deterministic differential
equation.

\bibliography{etinf.bib}

\end{document}